\documentclass{mem}
\usepackage{natbib}
\usepackage{txfonts}
\usepackage{balance}
\usepackage{graphicx}
\usepackage[a4paper]{hyperref}
\idline{0}{1}
\begin{document}
\def\teff{$T\rm_{eff }$}
\def\kms{$\mathrm {km s}^{-1}$}

\title{
TASTE: The Asiago Survey for Timing \\ transit variations of Exoplanets
}

   \subtitle{}

   \author{V. Nascimbeni\inst{1,2}\thanks{Visiting PhD Student at STScI under the DDRF D0001.82432
           program.} \and G. Piotto\inst{1} 
          \and L.\ R. Bedin\inst{2} \and M. Damasso\inst{1,3}
          }

   \institute{Dipartimento di Astronomia, Universit\`a degli Studi di Padova,
              Vicolo dell'Osservatorio 3, 35122 Padova, Italy \\
              \email{giampaolo.piotto@unipd.it, valerio.nascimbeni@unipd.it}
         \and
             Space Telescope Science Institute,
             3700 San Martin Drive, Baltimore, MD 21218 \\
             \email{bedin@stsci.edu}
         \and
             Astronomical Observatory of the Autonomous Region of the Aosta Valley, 
             Loc.\ Lignan 39, 11020 Nus (AO), Italy 
             \email{mario.damasso@studenti.unipd.it}
             }

\offprints{V. Nascimbeni}
 
\authorrunning{Nascimbeni et al.}

\titlerunning{the Asiago Survey for Timing transit variations of Exoplanets}

\abstract{ A promising method to detect earth-sized exoplanets is the
  timing analysis of a known transit. The technique allows  a search
  for variations in transit duration or center induced by the
  perturbation of a third body, e.g.  a second planet or an
  exomoon. To this aim, TASTE (The Asiago Survey for Timing
  transit variations of Exoplanets) project will collect
  high-precision, short-cadence light curves for a selected sample of
  transits by using imaging differential photometry at the Asiago
  1.82m telescope. The first light curves show that our project can already
  provide a competitive timing accuracy, as well as a  significant
  improvement over the orbital parameters. We derived refined ephemerides
  for HAT-P-3b and HAT-P-14b with only one transit each, thanks to a timing accuracy of 11 and 25 s, respectively.
}
\maketitle

\section{Introduction}

The transit of an exoplanet over the disk of its parent star
gives us the opportunity of a nearly complete characterization of 
system, by measuring orbital and physical parameters 
such as the orbital inclination $i$, 
the planetary radius $R_{\rm p}$, and hence the density of the planet.
Also, a focused follow-up of a transiting planet can reveal a second
planet in the system, following the \emph{Transit
Time Variation} (TTV) technique.  

\citet{holman2005} showed that the presence of a perturbing body 
(not necessarily transiting) breaks the strict periodicity of the 
primary planet, leading to predictable departures of 
the observed transit times from the computed mean ephemeris.
The amplitude of the TTV is  dependent upon the
mass of the perturber (more massive planets leading to a bigger
effect) and is strongly enhanced if the hidden planet is locked in a
low-order orbital resonance with the primary, like 1:2 or 2:3
\citep{agol2005}.  In that case, even a terrestrial planet could be
responsible for a TTV of the order of tens or hundreds of seconds.  
The \emph{Transit Duration Variation} (TDV) analysis is a
newer technique. A periodic  change
of the transit duration  may arise from the presence of a satellite,
or \emph{exomoon}, which makes the planet oscillate around the
planet-satellite barycenter along its path \citep{kipping2009a}.

Many projects about TTV/TDV analysis are currently ongoing,
the most notable being RISE
\citep{gibson2009} and the Transit Light
Curve Project \citep{holman2006}.  Most of these works  use an
instrumental setup which is specifically designed to get
high-precision light curves with a  short cadence --  sometimes a few
seconds. Systematic trends have to be
carefully identified and corrected, because they can 
mimick a fake TTV  \citep{pont2006,gibson2009}.  
Achieving a demonstrated timing accuracy $\lesssim$10 s is
still an ambitious target, typical values ranging from 15 to 30 s in
the best works.

Our primary aim is to collect a database of high-precision light
curves which will be suitable for a simultaneous TTV/TDV analysis,
optimizing every task from the observation/calibration setup to the
data extraction with fully home-made software tools, optimized for
this specific program. Our survey is based on data collected  with the
1.82m telescope at the
Asiago observatory\footnote{\sf http://www.pd.astro.it/asiago/}.
Some feasibility tests were also performed with the 0.81m telescope 
at the Osservatorio Astronomico della regione autonoma Valle d'Aosta\footnote{\sf
http://www.oavda.it/} (OAVdA) observatory.

\section{Methods}

\begin{figure}
\centering \includegraphics[width=6.7cm]{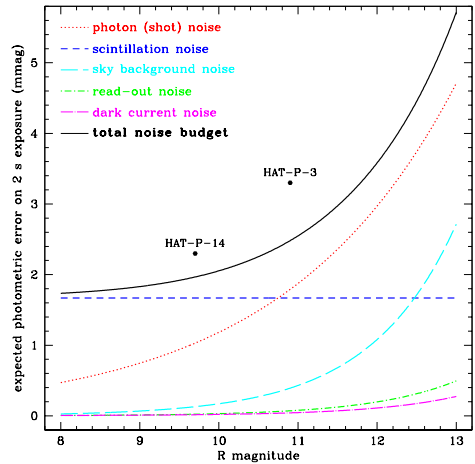}
\caption{Expected noise budget in mmag of a single 2 s exposure taken at the Asiago 1.82m 
telescope under typical observing conditions,
as function of the $R$ magnitude of the target. Measured off-transit scatters
are shown for the our light curves of HAT-P-14 and HAT-P-3.}
\label{nb}
\end{figure}

In the typical observing conditions we deal with (variable 
PSFs, lack of stellar crowding) the technique of
choice is differential aperture photometry, 
which allows us to normalize the flux of the target
with the flux of one or more
reference stars. The differential measure
automatically cancels out first order systematic trends, like
transparency variations. Bright targets are defocused to avoid
saturation and to minimize systematics 
from flat field residual errors and guiding drifts
\citep{southworth2009b}.  

\begin{figure*}[!t]
\centering 
\includegraphics[width=6.7cm]{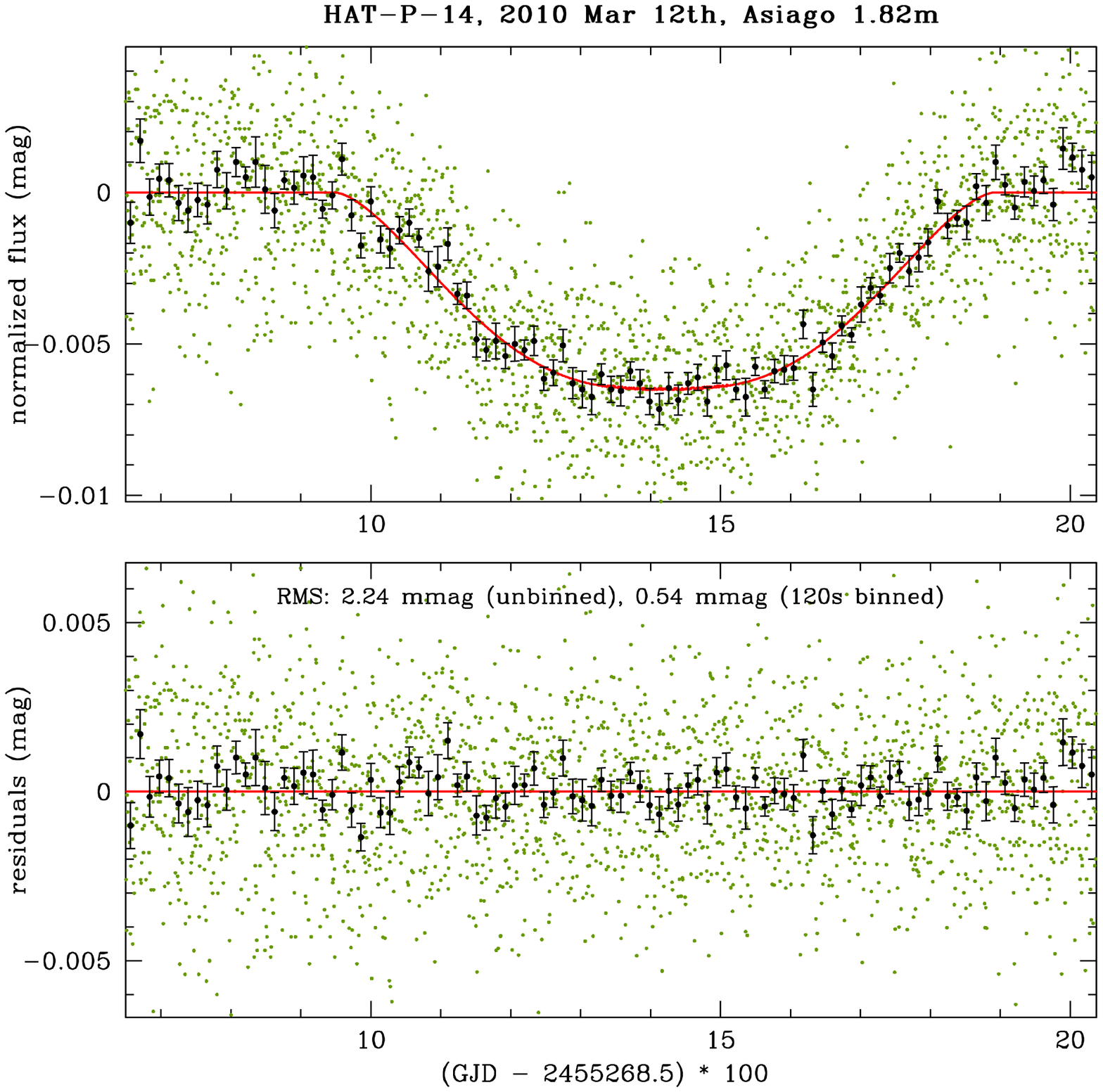}
\includegraphics[width=6.7cm]{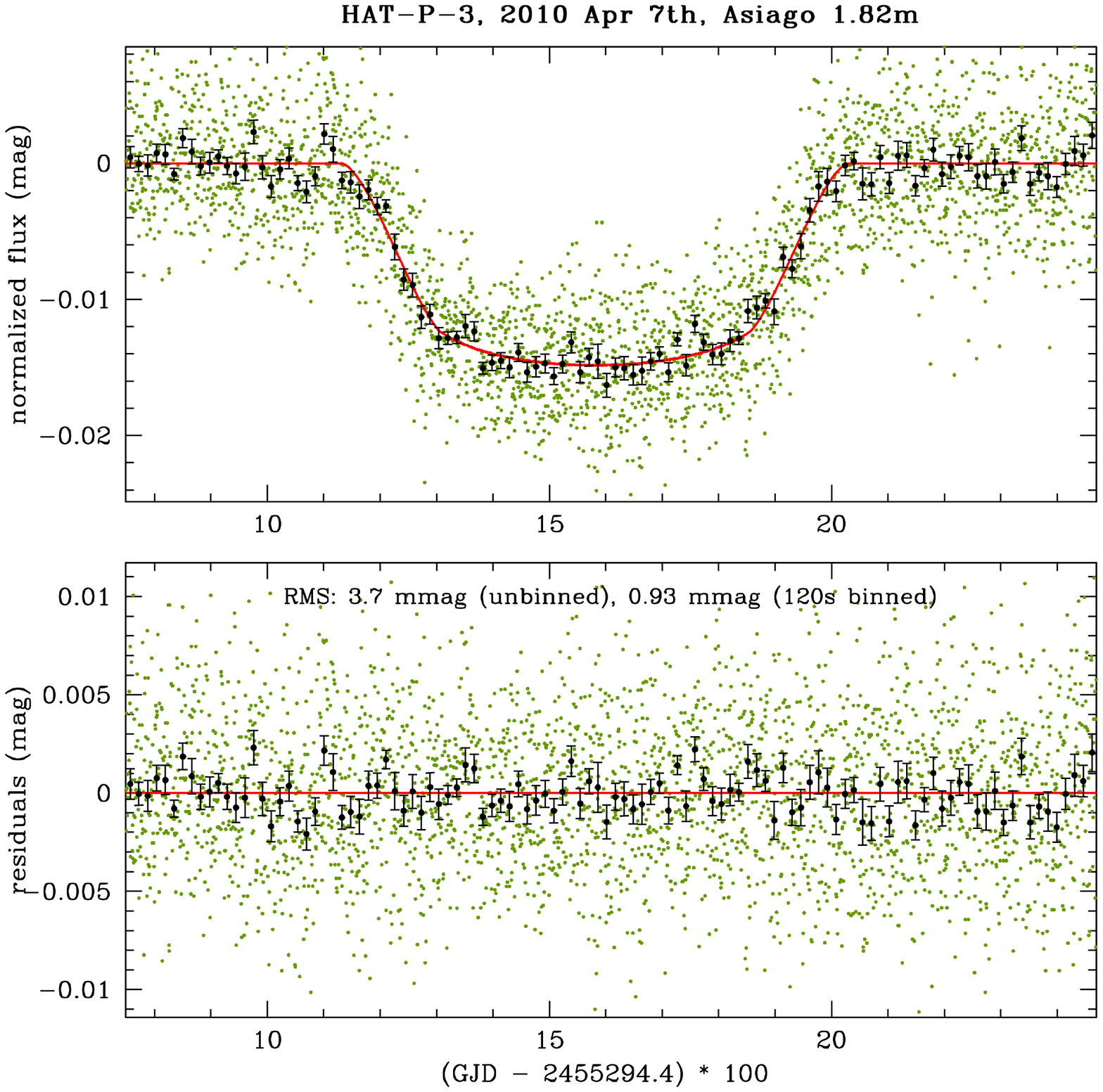}
\caption{
\emph{Left panel:} 
Light curve for HAT-P-14b 
($V= 9.98$, $\Delta V = 0.007$), observed on March 12, 2010 
with the Asiago 1.82m telescope, and residuals after the best fit model 
is subtracted. Unbinned points are shown in green and 120s-binned points in black.
Off-transit magnitude has been set to zero.
\emph{Right panel:} the same for the
light curve of HAT-P-3b ($V = 11.86$, $\Delta V = 0.013$), observed 
on April 7, 2010.
}
\label{lcs}
\end{figure*}

We employ as imager the Asiago Faint Object
Spectrograph and Camera (AFOSC), a focal-reducer type camera
with a 8.5$'$$\times$8.5$'$ field of view. 
AFOSC has been recently upgraded with a new E2V 42-20 thinned,
back-illuminated CCD (QE$\sim$90\% in the $R$ band). 
The tests we carried out
demonstrate that the detector is very stable, with low read-out noise
and dark current. The shutter is accurate down to 1 s
of exposure time. The camera controller software has been 
customized  to meet our needs for a very fast time sampling. 
4$\times$4 binning is used to lower the readout
time and decrease the readout noise. We read
a CCD sub-array  when only a part of the field is necessary. 
A big effort has been devoted to shorten as much as possible the dead
technical time between consecutive frames. 
An assessment of the expected S/N ratio shows that,
in typical conditions, our photometry is dominated by scintillation
for the brighter stars ($R\lesssim$11) and by photon noise 
for the fainter stars ($R\gtrsim$11). We expect for most of 
our targets a scatter $\lesssim$3 mmag on a single 2 s 
exposure (Fig. \ref{nb}).

In order to extract  the  maximal information from the available data, we
implemented and  use independent software tools  
specifically developed for this project.
The key concept of our pipeline is a fully empirical
approach to perform every task, including pre-processing, 
light curve extraction, and estimate of the transit
parameters along with their associated errors. The first half
of the pipeline, concerning the extraction of the light curves (STARSKY)
is already operative. The algorithms we have implemented are described in a 
forthcoming paper (Nascimbeni et al. submitted). 
We rely on the JKTEBOP code \citep{southworth2004}
to fit a model on our light curves. JKTEBOP includes routines for
the empirical estimation of the errors associated to the fitted parameters.

\section{First results}

\begin{figure*}[!t]
\centering 
\includegraphics[width=6.7cm]{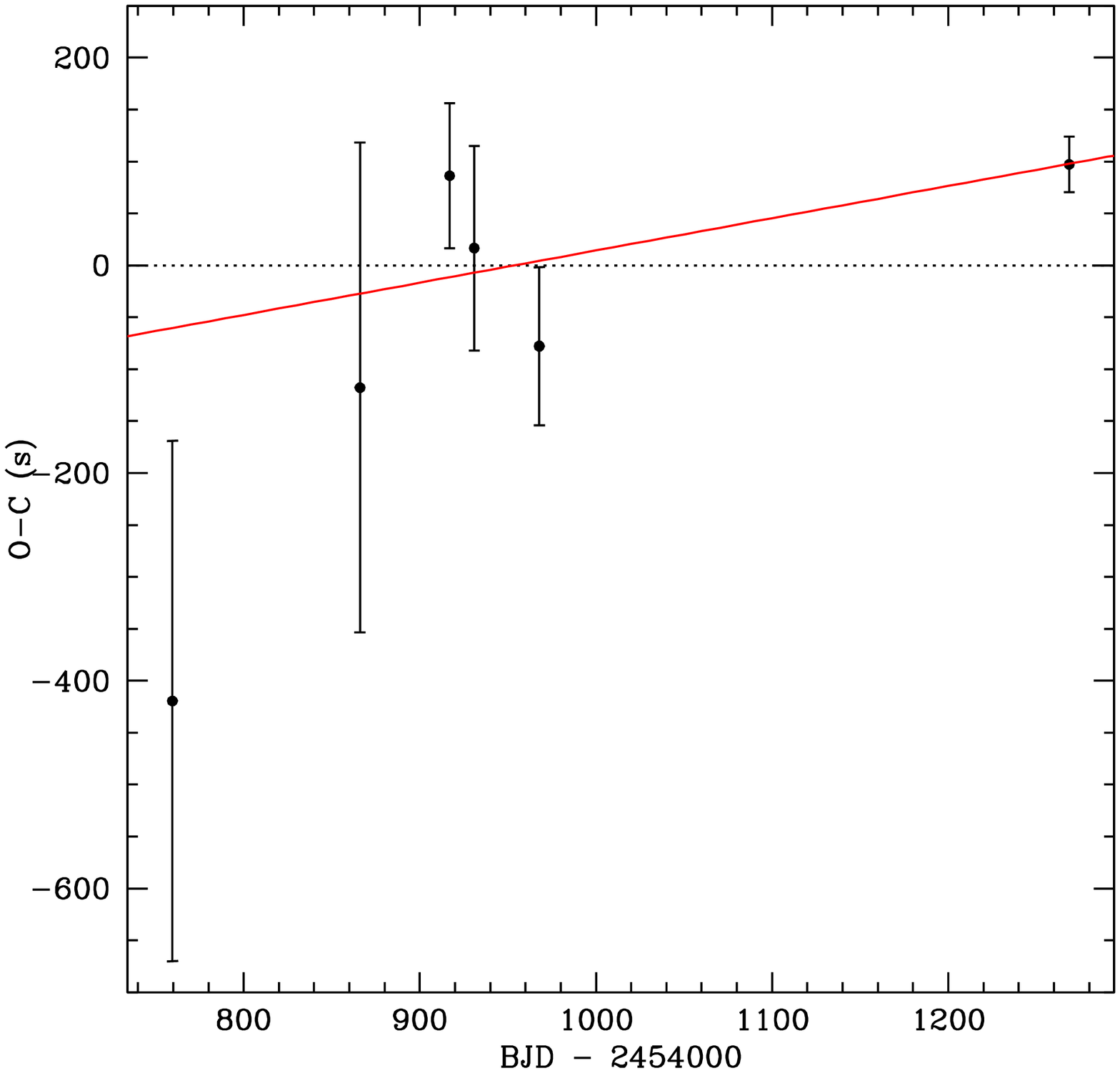}
\includegraphics[width=6.7cm]{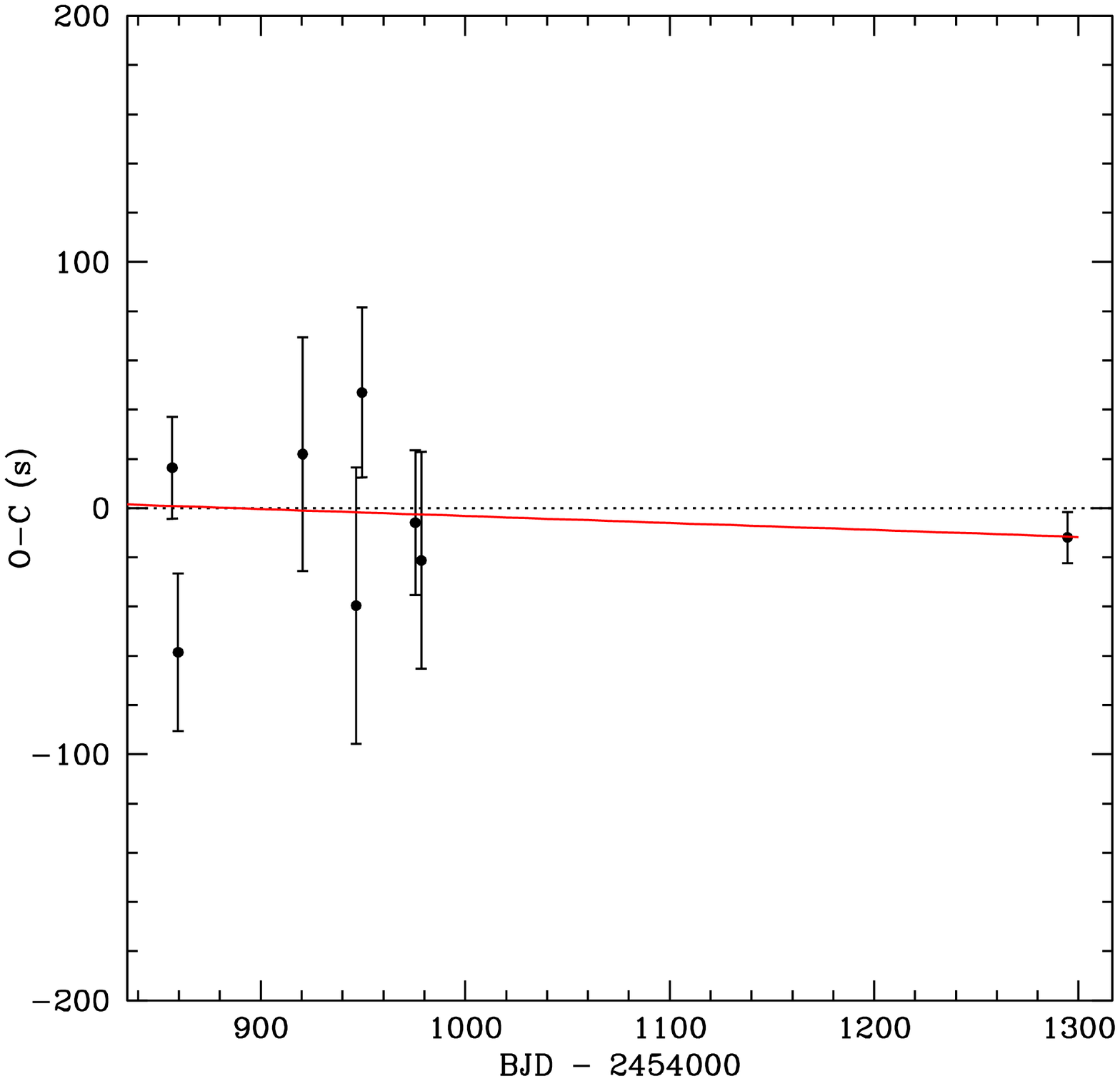}
\caption{$O-C$ diagram for HAT-P-14b (\emph{left panel}) 
and HAT-P-3b (\emph{right panel}). 
The last point in both plots is from TASTE; 
the others respectively from \citet{torres2010} and
\citet{gibson2010}. The red line is the weighted fit
for our refined ephemeris.}
\label{ttv}
\end{figure*}

Our survey already began to gather data. 
A sample of twelve targets has
been selected, based on both observational and astrophysical criteria. An
higher priority has been set for planets which would show the biggest
TDV/TDV signal by a 1 $M_\oplus$ perturber in a 2:1 resonant external orbit 
and by a 1 $M_\oplus$ exomoon at one third of the Hill radius. Targets
such as WASP-3b were included because of an existing TTV claim (Maciejewski 
et al. 2010).

We report the first collected light curves for HAT-P-14b and HAT-P-3b
\citep{torres2010,torres2007}, 
which are respectively the brightest, and near to the faintest magnitude 
limits of our surveyed sample (Fig \ref{lcs}). The unbinned series is
made from 2,247 and 2,882 frames respectively, with a net
time sampling of 5.4 and 5 s and an RMS 
of 2.2 and 3.7 mmag around the best-fit model.
The overall photometric performances of our system met our
expectations. The scatter measured on our
light curves is in good agreement with the theoretical expected amount of
noise, as shown in Fig. \ref{nb}. 
In particular, the noise measured for HAT-P-14
($\sim 0.54$ mmag on 120 s bins) indicates a very low amount of
systematic errors, and is of the same order achieved by
state-of-the-art photometry on medium class telescopes 
(e.g., see Southworth et al. 2009b, with data acquired in a much better site).

The timing accuracy achieved by our observations was assessed by
different techniques. The most conservative one (a ``prayer-bead'' residual
permutation algorithm) returned for the central instant $T_0$ an error of
$\sim$25 s for HAT-P-14b and $\sim$11 s for HAT-P-3b. The improvement
over the previous measurements is made clear by comparing their errorbars
with ours in the Observed $-$ Calculated diagram for $T_0$ (Fig. \ref{ttv}).
For both planets, we were able to refine the ephemeris and the orbital 
parameters with only one observed transit. 

\section{Conclusions}

The performance achieved by TASTE photometry proved it to be suitable
for a long-term characterization of the targets and for an extremely
accurate TTV/TDV analysis. Further improvements are still possibile for the data 
analysis and for the duty-cycle of our series. Once a sufficient number of light curves
will be collected for a given target, it should be possible to search
for low-mass perturbers, and to put very stringent upper-limits
in case of null detection.

\begin{acknowledgements}
We thank M. Fiaschi for the support in the optimization for our
purposes of AFOSC data acquisition software. 
\end{acknowledgements}

\bibliographystyle{aa}
\bibliography{biblio.bib}

\end{document}